\documentstyle[12pt]{article}
\begin{document}

\title{Reason for the large scale differences in the results for the
neutron-antineutron oscillation problem}
\author{V.I.Nazaruk\\
Institute for Nuclear Research of RAS, 60th October\\
Anniversary Prospect 7a, 117312 Moscow, Russia.\\
E-mail: nazaruk@al20.inr.troitsk.ru}

\date{}
\maketitle
\bigskip

\begin{abstract}

It is shown that there is a double counting in the standard model of $n\bar
{n}$ mixing in the medium, resulting in full cancellation of leading terms.
The direct calculation of $n\bar{n}$ transition followed by annihilation is 
performed. The lower limit for the free-space $n\bar{n}$ oscillation time is 
$\tau_{n\bar{n}} \sim T_{n\bar{n}}>10^{16}y$, where $T_{n\bar{n}}$ is the 
lifetime of neutron bound in a nucleus. This limit exceeds the previous one
by 16 orders of magnitude.
\end{abstract}

\vspace{5mm}

\newpage
\setcounter{equation}{0}

Any information on the occurrence of $n\bar{n}$ oscillation[1,2] is important 
in order to discriminate among various grand unified theories. The most direct
limit on the free-space $n\bar{n}$ oscillation time $\tau_{n\bar{n}}$ is 
obtained using free neutrons:\ $\tau_{n\bar{n}}>10^{7}s$[3]. Alternatively, a
limit can be extracted from the nuclear annihilation lifetime measured in 
proton-decay type experiments:\ $\tau_{n\bar{n}}>10^{8}s\sim 1y$ (see, for 
example, Ref.[4,5]). The process under consideration is $(nucleus)\rightarrow 
(\bar{n}-nucleus)\rightarrow (annihilation\; products)$. The calculations were 
based on the potential model of $n\bar{n}$ mixing in the medium [4] or on 
nonrelativistic diagram technique [5]. They predict the drastical suppression 
of $n\bar{n}$ transition in nuclei. However, $n\bar{n}$ conversion comes from 
the exchange of Higgs with $m_H>10^5GeV$, so from the point of view of the 
microscopic theory (dynamic $n\bar{n}$ conversion[2], annihilation) the 
reasons for the suppression are not clear. 

In this paper it is shown that the models used previously [4,5] are too crude
or, more categorically, inapplicable to the problem under study. In particular,
the potential model does not correspond to annihilation products in the final 
state. We perform the direct calculation of the process $(nucleus)\rightarrow 
(\bar{n}-nucleus)\rightarrow (annihilation\; products)$. The interaction 
Hamiltonian is taken in the general form. Then the basic part of calculation 
is a model-independent. However, the amplitude in this case is singular. For 
solving the problem the approach with finite time interval [6] is used.

{\bf 1.}In the standard approach (later on reffered to as potential model) the 
$n\bar{n}$ transitions in the medium are described by Schrodinger equations
\begin{eqnarray}
(i\partial_t+\nabla^2/2m-U_n)n(x)=\epsilon \bar{n}(x),\nonumber\\
(i\partial_t+\nabla^2/2m-U_{\bar{n}})\bar{n}(x)=\epsilon n(x).
\end{eqnarray}
Here $\epsilon=1/\tau_{n\bar{n}}$ is a small parameter[4]; $U_n$ and
$U_{\bar{n}}$ are the self-consistent neutron potential and the 
$\bar{n}$-nucleus
optical potential respectively. For $U_n=const.$ and $U_{\bar{n}}=const.$ in
the lowest order in $\epsilon$ the probability of the process is
\begin{eqnarray}
W_{pot}(t)=1-\mid U_{ii}(t)\mid ^2=2ImT_{ii}(t),\nonumber\\
T_{ii}(t)=i(\epsilon/\delta U)^2[1-i\delta Ut-\exp (-i\delta Ut)],
\end{eqnarray}
where $U$ and $T$ are the evolution operator and $T$-operator 
respectively; $U=1+iT$ and
\begin{equation}
\delta U=U_{\bar{n}}-U_n=ReU_{\bar{n}}-i\Gamma /2-U_n.
\end{equation}
Here $\Gamma\sim 100 MeV$ is the annihilation width of $\bar{n}$-nucleus state.

What is described by $W_{pot}(t)$? Let us take the imaginary part of Eq.(16) of
Ref.[6]
\begin{equation}
2ImT_{ii}(t)=\epsilon^2t^2-\epsilon^2\int_0^tdt_{\alpha}\int_0^
{t_{\alpha}}dt_{\beta}2ImT_{ii}^{\bar{n}}(\tau ),
\end{equation}
where $\tau=t_{\alpha}-t_{\beta}$. Here $T(t)\;and\;T^{\bar{n}}(\tau )$ are 
the $T$-operators of the whole process and $\bar{n}$-nucleus interaction 
respectively. For $T^{\bar{n}}$-operator $\mid\!i\!>=\mid\!0\bar{n}_p\!> $ and 
$<\!f\!\mid $ are the $\bar{n}$-nucleus and annihilation products respectively.
For the whole $T$-operator $\mid\!i\!>=\mid\!0n_p\!>$ is the nucleus, the 
physical meaning of the final states will be cleared up later. The probability 
conservation $\sum_{f}\mid U_{fi}\mid ^2=1$ gives us
\begin{eqnarray}
2ImT_{ii}=\sum_{f\neq i}\mid T_{fi}\mid ^2+\mid T_{ii}\mid ^2,\nonumber\\
\sum_{f\neq i}\mid T_{fi}(t)\mid ^2=W(t).
\end{eqnarray}
The probability of the process $W(t)$ will be defined below;  $\mid T_{fi}\mid 
^2\sim \epsilon^2$, whereas $\mid T_{ii}\mid ^2\sim\epsilon^4$ (see Eqs.(2) and
(15)). So for the l.h.s. of Eq.(4) $2ImT_{ii}(t)=W(t)$, that was taken into 
account in (2). For the $T$-matrix of $T^{\bar{n}}(\tau )$-operator Eq.(5) has 
the form
\begin{equation}
2ImT_{ii}^{\bar{n}}(\tau)=\sum_{f\neq i}\mid T_{fi}^{\bar{n}}
(\tau )\mid ^2+\mid T_{ii}^{\bar{n}}(\tau )\mid ^2,
\end{equation}
$T_{ii}^{\bar{n}}=<\!0\bar{n}_p\!\mid T^{\bar{n}}\mid \!0\bar{n}_p\!>$,
$T_{fi}^{\bar{n}}=<\!f\!\mid T^{\bar{n}}\mid \!0\bar{n}_p\!>$.
The $\bar{n}$-nucleus interaction is nonperturbative process and $\mid T_{ii}^
{\bar{n}}\mid ^2\sim\sum_{f\neq i}\mid T_{fi}^{\bar{n}}\mid ^2 $. Now Eq.(4) is
\begin{equation}
W(t)=\epsilon^2t^2-\epsilon^2\!\int_0^t\!dt_{\alpha}\!\int_0^{t_
{\alpha}}\!dt_{\beta}\mid <\!0\bar{n}_p\!\mid T^{\bar{n}}(\tau )\mid \!0
\bar{n}_p\!> \mid ^2-\epsilon^2\!\int_0^t\!dt_{\alpha}\!\int_0^{t_{\alpha}}\!
dt_{\beta}\sum_{f\neq i}\mid <\!f\!\mid T^{\bar{n}}(\tau )\mid \!0\bar{n}_p\!>
\mid ^2.
\end{equation}

Let us calculate $T_{ii}^{\bar{n}}$ and $T_{fi}^{\bar{n}}$ in the framework
of potential model. The wave function of initial state is described by the 
equation
\begin{equation}
i\frac{\partial \Phi }{\partial t}=H_0\Phi,
\end{equation}
$H_0=-\nabla^2/2m+U_n$. At the moment $t=0$ the interaction $\delta U$ is 
turned on. We have
\begin{equation}
i\frac{\partial \Psi }{\partial t}=(H_0+\delta U)\Psi ,
\end{equation}
$\Psi(0)=\Phi (0)$. The projection to the initial state and $T$-matrix at 
$t=\tau $ are 
\begin{eqnarray}
<\Phi \mid \Psi >=U_{ii}^{\bar{n}}(\tau)=\exp (-i\delta U\tau),\\
T_{ii}^{\bar{n}}(\tau )=i[1-\exp (-i\delta U\tau )],\nonumber\\
\sum_{f\neq i}\mid T_{fi}^{\bar{n}}(\tau )\mid ^2=1-\mid U_{ii}^{\bar{n}}
(\tau)\mid ^2=1-e^{-\Gamma\tau}=W_{\bar{n}}(\tau),
\end{eqnarray}
where $W_{\bar{n}}(\tau)$ is the $\bar{n}$-nucleus decay probability. Note that
$\Gamma $ corresponds to all $\bar{n}$-nucleus interactions followed by
annihilation. However, the main contribution comes from the annihilation 
without rescattering of $\bar{n}$[5], because $\sigma _{ann}>2\sigma _{sc}$.
Substituting these expressions in (7), one obtains the potential model result
(2).

Therefore, the approach with the finite time interval was verified by the 
example of exactly solvable potential model. It is involved in Eq.(4) as a 
special case. Solving Eqs.(1) by method of Green functions we will obtain the 
same results. We have started from Eq.(4) only for verification of the finite 
time approach. 

Let us return to Eq.(7). It is at least unclear. (1)The first term is 
free-space $n\bar{n}$ transition probability. The matrix elements $T_{ii}^{\bar
{n}}$ and $T_{f\neq i}^{\bar{n}}$ describe transitions $(\bar{n}-nucleus)
\rightarrow (\bar{n}-nucleus)$ and $(\bar{n}-nucleus)\rightarrow
(annihilation\; products)$ respectively. So the first and the second terms
corresponds to $\bar{n}$-nucleus in the final states. However, in the
experiment only the annihilation products are detected ($\bar{n}$-nucleus is
unobservable) and the result should be expressed only in the terms of  $T_
{f\neq i}^{\bar{n}}$. Moreover, $\bar{n}$-nucleus decays into final states 
identical with the states given by the third term. This suggests that the 
potential model contains the double counting. Expression $1-\mid U_{ii}\mid 
^2$ from Eq.(2) describes the inclusive decay of initial state and so the $n
\bar{n}$ transition with $\bar{n}$-nucleus in the final state is also included 
in $W_{pot}$, unless additional limits are imposed. To exclude the double 
counting the annihilation products in the final state should be fixed. (2)Let 
us $\mid \delta Ut\mid \ll 1$. (This is the case in some other problems.) When 
$\Gamma =0$, the third term equals to zero. When $\Gamma \neq 0$, the 
contribution of the third term is negative and $dW/d\Gamma <0$, whereas the 
opening of the new channel (annihilation) should increase $W$.

How big is the probable error? The contributions of the second and third terms
are: $x_2=-\epsilon^2t^2/2+F_2$, $x_3=-\epsilon^2t^2/2+F_3$. The functions 
$F_{2,3}$ contain the terms proportional to $t$ and $\exp (-i\delta Ut)$. 
So the $\epsilon^2t^2$ term produced by the third term is fully canceled. This 
is a consequence of double counting. Therein lies the reason of the 
discrepancy between our result and the result of the potential model. 

As noted in[6], Eqs.(11) and (2) can also be obtained by means of microscopic
variant of the potential model (zero angle rescattering diagrams of $\bar{n}$).
In this case the Hamiltonian of $\bar{n}$-medium interaction is $H=\delta U$. 
The same calculation was repeated by Dover et al.[4]. They substitute $H=-i
\Gamma /2$ in (4) and obtaine (2). On the basis of this and only this they 
refute the result of Ref.[6]. In other words they refute our limit because it 
differs from the prediction of the potential model ($H=-i\Gamma /2$[4]).

What is wanted is $\sum_{f\neq i}\mid T_{fi}\mid ^2$, where $<\!f\!\mid$ is 
the annihilation products. It is connected with the diagonal matrix element 
by Eq.(5):
\begin{equation}
2ImT_{ii}=\sum_{f\neq i}T_{fi}^*T_{fi}.
\end{equation}
Calculation of $T_{ii}$ is determined by r.h.s. of Eq.(12): the cut
corresponding to $T_{ii}$ must contain only annihilation products, that is not
in accordance with Eq.(7). It includes redundent states $f=(\bar{n}-nucleus)$
forbidden by the unitarity condition. The relation (12) is not fulfilled. Also 
the eigenfunctions of $H_0+\delta U$ do not form the complete orthogonal set. 
Due to this fact the $\bar{n}$-nucleus (described by $U_{\bar{n}}$) also can 
not appear in Eq.(12) as the intermediate state. So the model (1) is 
inapplicable in our case because it leads automatically to incorrect matrix 
element $T_{ii}$. Elimination of redundent trajectories from $T_{ii}$ means 
the direct calculation of $T_{fi}$. 

{\bf 2.}In Ref.[6] the first and the third terms were taken into account. The
second one was omitted. The first term corresponds to low density limit and is
meaningfull for $n\bar{n}$ transitions in the gas. This scheme is not quite 
correct here. In this paper we perform the direct calculation of the process 
$(nucleus)\rightarrow (\bar{n}-nucleus) \rightarrow (annihilation\; products)$.
We have
\begin{equation}
<f\mid U(t,0)-I\mid0n_p>=iT_{fi}(t)=
\sum_{k=1}^{\infty}(-i)^{k+1}<f\mid \int_0^tdt_1...\int_{0}^{t_{k-1}}dt_k
\int_{0}^{t_k}dt_{\beta }H(t_1)...H(t_k)H_{n\bar{n}}
(t_{\beta })\mid\!0n_p\!>,
\end{equation}
where
\begin{eqnarray}
H(t)=(all\;\; \bar{n}-medium\;\; interactions) - U_n,\nonumber\\
H_{n\bar{n}}(t)=\epsilon \int d^3x(\bar{\Psi }_{\bar{n}}\Psi _n+h.c.),
\end{eqnarray}
$H+H_{n\bar{n}}=H_I$. Here $\mid \!0n_p\!>$ is the state of the medium 
containing the neutron with 4-momenta $p=({\bf p}_n^2/2m+U_n,{\bf p}_n)$, 
$<\!f\!\mid$ represents the annihilation products; $H_{n\bar{n}}$ is the 
oscillation Hamiltonian[4]. In the case of the formulation of the $S$-matrix 
problem $(t,0)\rightarrow (\infty ,-\infty)$ Eq.(13) in the momentum 
representation includes the singular propagator $G=1/(\epsilon_n-{\bf p}^2_n/
2m-U_n)\sim 1/0$. Taking into account that $H_{n\bar{n}}\mid\!0n_p\!>=\epsilon 
\mid\!0\bar{n}_p\!>$, we change the order of integration and obtain
\begin{eqnarray}
T_{fi}(t)=- \epsilon \int_{0}^{t}dt_{\beta }iT_{fi}^{\bar{n}}(t-t_{\beta }),
\nonumber\\
iT_{fi}^{\bar{n}}(\tau )=\sum_{k=1}^{\infty}(-i)^k\int_
{t_{\beta }}^{t}dt_1...\int_{t_{\beta }}^{t_{k-1}}dt_k<f\mid H(t_1)...H(t_k)
\mid\!0\bar{n}_p\!>,
\end{eqnarray}
where $\mid\!0\bar{n}_p\!>$ is the state of the medium containing the $\bar
{n}$ with 4-momenta $p$; $\tau=t-t_{\beta}$. The 4-momenta of $n$ and $\bar
{n}$ are equal. $T_{fi}^{\bar{n}}$ is an exact amplitude of $\bar{n}$-nucleus 
decay. It includes all the $\bar{n}$-nucleus interactions followed by 
annihilation. Expression for $T_{fi}(t)$ was obtained in perfect analogy to 
(4) that can be considered as a test for Eq.(15).

The 2-step process was reduced to the annihilation decay of $\bar{n}$-nucleus.
(The slightly different method is the separation of the antineutron Green 
function [6].) It is seen from (13) and (15) that both pre- and post- $n\bar
{n}$ conversion spatial wave functions of the system coincide:
\begin{equation}
\label{16}\mid\!0n_p\!>_{sp}=\mid\!0\bar{n}_p\!>_{sp}.
\end{equation}
$\bar{n}$ appears in the state with $\delta U=0$. We would like to stress that 
in the potential model (1) the picture of $\bar{n}$-nucleus formation is 
exactely the same: in Eq.(4) for $T_{ii}=<\!0n_p\!\mid T\mid\! 0n_p\!>$ and 
$T_{ii}^{\bar{n}}=<\!0\bar{n}_p\!\mid T^{\bar{n}}\mid\!0\bar{n}_p\!>$ 
condition (16) was fulfilled. Hereafter, the potential model of the $\bar{n}$-
medium interaction (block $T^{\bar{n}}$) was used and $W_{pot}$ was 
reproduced, which confirms the picture of $\bar{n}$-nucleus formation given 
above. Solving Eqs.(1) by method of Green functions we will obtain the same 
results, including (16). The equality of vectors of state (16) is also evident 
from the continuity of solution of Eqs.(1).

In both models the first stage of the process ($n\bar{n}$ conversion) is 
described identically. The basic difference centers on the next stage - 
annihilation. In 
the potential model $T_{ii}^{\bar{n}}$ is calculated (as a result the 
self-energy part $\Sigma =\delta U$ appears) and is used in Eq.(7), which is 
wrong. We calculate $T_{fi}^{\bar{n}}$ starting from the same point (16). The 
result will be expressed through $\Gamma $ (see Eqs.(19),(11)), but not through
$\delta U=Re\delta U-i\Gamma /2$, as is usually the case in decay calculations.
The standard $\delta U$-dependence is manifested in scattering problems, when 
the diagonal matrix element  $T_{ii}$ in r.h.s. of Eq.(12) should be taken into
account. It corresponds to the observarble process - zero angle scattering of 
the incident particle.

The characteristic annihilation time of $\bar{n}$ is: $\Delta =1/\Gamma \sim 
10^{-23}$s. When $\tau \gg \Delta$, $T_{fi}^{\bar{n}}(\tau )$ reaches its 
asimptotic value $T_{fi}^{\bar{n}}$:
\begin{equation}
\label{17}T_{fi}^{\bar{n}}(\tau \gg \Delta )=T_{fi}^{\bar{n}}(\infty )=
T_{fi}^{\bar{n}}=const.
\end{equation}
The expressions of this type are the basis for all $S$-matrix calculations.
(Measurement of any process corresponds to some interval $\tau $. So it is 
necessary to calculate $U(\tau )$. The replacement $U(\tau )\rightarrow 
S(\infty )$ is equivalent to (17).) Let us take $t\gg \Delta$. From (15) and 
(17) we have
\begin{equation}
\label{18}T_{fi}(t)=-i\epsilon [\int_{0}^{t-\Delta }dt_{\beta }T_{fi}^{\bar
{n}}(t-t_{\beta })+\int_{t-\Delta }^{t}dt_{\beta }T_{fi}^{\bar{n}}(t-t_
{\beta })]\sim -i\epsilon tT_{fi}^{\bar{n}}.
\end{equation}
The contribution of the second term is negligible since $\mid T_{fi}^{\bar{n}}
(\tau)\mid ^2\leq 1$. The probability of the whole process is
\begin{equation}
\label{19}W(t)=\sum_{f\neq i}\mid T_{fi}(t)\mid ^2\sim \epsilon^2t^2
 \sum_{f\neq i}\mid T_{fi}^{\bar{n}}(t)\mid ^2\sim \epsilon^2t^2,
\end{equation}
where Eq.(11) has been taken into account. The value $\epsilon^2t^2=t^2/\tau_
{n\bar{n}}^2$ is the free-space $n\bar{n}$ transition probability. Due to the 
annihilation channel $n\bar{n}$ conversion is practically unaffected by the 
medium. So $\tau_{n\bar{n}}\sim T_{n\bar{n}}$, where $T_{n\bar{n}}$ is the 
oscillation time of
neutron bound in a nucleus. In order to find the limit for $\tau_{n\bar{n}}$ 
from experimental data on nuclear stability, the distribution (19) should be 
used (but not the exponential decay law!). Let us $N_n,T_0,\epsilon _1$ and 
$\theta $ are the total number of neutrons under observation, the observation 
time, the overal $n\rightarrow \bar{n}$ detection efficiency and the average 
number of observable $n\rightarrow \bar{n}$ events respecrively. From the 
inequality
\begin{equation}
\label{20}N_n(T_0/\tau_{n\bar{n}})^2(\epsilon _1/\theta )<1
\end{equation}
one obtains $\tau_{n\bar{n}}>10^{16}y$, where the values $T_0=1.3y$, 
$N_n=2.4\cdot 10^{32}$, $\epsilon _1=0.33$ and $\theta =2.3$ [7] were used.

Our previous result [6] is different from (19) only by a factor of 2. However,
in Ref.~[6] we used the limit $T_{n\bar{n}}>4.3\cdot10^{31}y$ [7] deduced from 
the experimental data by means of exponential decay law which does not agree 
with (19).

{\bf 3}.Let us return to the reason of enormous quantitative disagreement
between our result and the potential model one. The strong sensitivity of the
results should be expected. Really, in the momentum representation the 
$S$-matrix amplitude $M_s$, corresponding to $n\bar{n}$ transition followed by
annihilation (see Eq.(14)) diverges
\begin{equation}
\label{21}M_s=\epsilon\frac{1}{\epsilon_n-{\bf p}^2_n/2m-U_n}M\sim \frac{1}{0},
\end{equation}
where $M$ is the annihilation amplitude. These are infrared singularities
conditioned by zero momentum transfer in the $\epsilon $-vertex. It is can
be seen that $M_s\sim 1/0$ for any bound state wave function of a neutron
(i.e., for any nuclear model). On the other hand in the potential model the 
energy is not conserved and becomes complex: $M_A\rightarrow M_A+\delta U$ 
($M_A$ is the nuclear mass). In this case we have instead of (21) 
\begin{equation}
\label{22}M_{pot}=\epsilon\frac{1}{\epsilon_n-{\bf p}^2_n/2m-U_{\bar{n}}}M=
\epsilon\frac{1}{\delta U}M.
\end{equation}
This is a potential model amplitude. Really, the process width is $\Gamma 
_{pot}=\int d\Phi \mid M_{pot}\mid ^2/2M_A=\epsilon ^2\Gamma /\mid \delta U
\mid ^2=W_{pot}/t$, that coincides with (2) when $\mid \delta Ut\mid \gg 1$. 
It is seen that: (1) There is a double counting in $M$ and $G$ with respect to 
$H$. $M_{pot}$ does not agree with Eq.(14) as well. (2) $\delta U=0$ is the 
singular point and due to zero momentum transfer $q=0$ in the vertex 
corresponding to $H_{n\bar{n}}$ we are in this point. So the result is 
extremely sensitive to $\delta U$. (Usually, in the reactions and decays the 
momentum transferred is $q\neq 0$. In this case $\delta U$-dependence of $G$ 
is masked by  $q$: $G^{-1}=(\epsilon_n-q_0)-({\bf p}_n-{\bf q})^2/2m-U_n-
\delta U$. We deal with the 2-tail and $q=0$.)

Comparing (21) with (18) one sees that in principle the limit $\delta U
\rightarrow 0$ corresponds to the replacement
\begin{equation}
\label{23}1/\delta U\rightarrow t.
\end{equation}
Certainly, we do not set $\delta U=0$. $U_{\bar{n}}$ is not introduced at all. 
In the calculation of Eq.(13) the multiplier $t$ (see Eq.(18)) arises 
automatically instead of $1/\delta U$ in the potential model, or $1/\Delta q$ 
in the case $q\neq 0$. When $q\neq 0$ in the $\epsilon$-vertex, Eq.(13) leads 
to usual $S$-matrix result (see below). Formal reason for the differences in 
the results is the full cancellation of the terms $\sim t^2$ in Eq.(7). 
Erroneous structure of (7) is caused by nonperturbative and 2-step nature of 
the process. The fact that $q=0$ extremely increases the disagreement.

{\bf 4}.An additional comment is necessary regarding t-dependence of the whole
process probability $W(t)$. Eq.(19) has been obtained in the lowest
order in $\epsilon$. The exact distribution $W_{pr}(t)$ which accounts for
the all orders in $\epsilon$ is unknown. However, $W$ is the first term of the
expansion of $W_{pr}$ and we can restrict ourselves to the lowest order
$W_{pr}=W$, as it is usully the case for rare decays. $W_{pot}$ is also 
calculated in the lowest order in $\epsilon$.

The protons must be in very early stage of the decay process. Thus the
realistic possibility is considered [8-11] that the proton has not yet entered
the exponential stage of its decay but is, instead, subject to non-exponential
behavior which is rigorously demanded by quantum theory for sufficiently early
times. At first sight, since $\tau_{n\bar{n}}>10^{16}y$ for $n\bar{n}$-mixing
in a nuclear, the non-exponential behavior should be expected too. In fact, 
there is one more problem: we deal with the two-step process. When trying to 
calculate $M_s$ and $\Gamma _s$ in the framework of 
standard S-matrix theory we get $\Gamma _s\sim 1/0$. So the decay law 
$\exp (-\Gamma _st)$ is unrelevant and it is necessary to deduce the
distribution $W(t)$ as it was done above.

We have to mention the main points of Krivoruchenko's preprint[12]. (1)The 
$n\bar{n}$ transition followed by annihilation (two-step nuclear decay) and 
motion of particle in the classical field are two different problems.
Describing the first one by Eqs.(1) we understand that this is an effective
procedure. From formal standpoint in the first and the second cases the 
potentials are complex and real respectively. Unfortunately, sometimes the 
literal analogy between these problems is drawn[12]. (2)The initial Eq.(11) of 
Ref.[12] must describe the $n\bar{n}$ transition followed by annihilation. 
However, the l.h.s. of Eq.(11) is free of $\bar{n}$-nucleus interaction at all.
The r.h.s. contains annihilation width $\Gamma$ (we stress this point) and 
coincides with the potential model result. We also would be glad to get the 
result without calculations, but some difficulties emerge in reaching this 
goal.

The interaction responsible for the $n\bar{n}$ conversion is ultra-weak.
Therefore, the $n$-nucleus interaction in the initial state should be taken
into account exactly. The neutron line entering into the $n\bar{n}$ transition 
vertex should be the wave function of the bound state (see Eq.(8)), but not 
the propagator, 
as in the model based on diagram technique[5,13]. As a result, in this model 
the $n\bar{n}$ transition is possible only between the acts of interactions of 
oscillating particle and a nucleus. These interactions lead to total 
suppresion of $n\bar{n}$ conversion, that is incorrect. This can be understood 
using the analogy with $\beta $ decay and taking into account that $m_H>10^5$
GeV. 

Some additional remarks to paper[13] are necessary. (1)The picture described 
in Sec.1 is valid only for simple interaction operator. We deal with products 
of operators (see Eq.(13) of this paper). (2) In Sec.2 it is claimed that the 
amplitude should be singular (!) at $B\rightarrow 0$, where $B$ is the binding
energy. Accordingly, as $B\rightarrow 0$, the amplitude obtained $\mid A_1\mid 
^2\sim 1/0$. In fact, they are usual infrared singularities mentioned above,
wich must be avoided. The correct model should reproduce the law dencity limit 
$W(t)=\epsilon ^2t^2$. (3)The cut corresponding to the diagonal matrix element 
(18) is completely free of annihilation products.

We try to calculate the process amplitude starting from (14). The $S$-matrix 
theory gives (21). The approach with finite time interval is infrared-free. 
Its verification for the diagrams with $q=0$ was made above by the example
of potential model. For nonsingular diagrams the test is obvious. Let us $q
\neq 0$ in the $\epsilon$-vertex. The appropriate calculations with finite time
interval (adiabatic hypothesis should be used) give the $S$-matrix result (we 
stress this fact because it means the verification of the approach):  $T_{fi}=
i\epsilon '(1/\Delta q){\cal T}^{\bar{n}}_{fi}$, where ${\cal T}^{\bar{n}}_
{fi}$ is the $S$-matrix amplitude of annihilation of virtual $\bar{n}$ with 
4-momenta $k=p-q$. Comparing with (18) one sees that limit $\Delta q
\rightarrow 0$ corresponds to the replacement $1/\Delta q\rightarrow t, {\cal 
T}^{\bar{n}}_{fi}\rightarrow T_{fi}^{\bar{n}}$ (compare with (23)). The similar
problem for matrix element $T_{ii}$ was solved in Ref.[14]. 

The main results of this paper are given in the abstract. In the next paper
the following statements will be proved.(1) All the results are true for any
nuclear model.(2) The contribution of the corrections is negligible. (3) 
Further investigation and verification of the approach will be presented as
well. In our opinion, it makes sense to look at some other problems on 
oscillation of particles in a medium from the standpoint given above.

\end{document}